# High fidelity simulations of the multi-species Vlasov-Maxwell system with the Numerical Flow Iteration

R.-Paul Wilhelm[1] and Fabio Bacchini[1,2]

[1]Centre for mathematical Plasma Astrophysics (CmPA), KU Leuven, Belgium
[2]Royal Belgian Institute for Space Aeronomy, Brussels, Belgium

E-mail: paul.wilhelm@kuleuven.be

**Abstract.** Validity of fluid models breaks down for non-thermal or weakly collisional plasmas which often occur e.g. in the solar wind. In these regimes one has to resort to modelling through the first-principle Vlasov–Maxwell system, but its six-dimensional phase-space dynamics, strong filamentation, and multi-scale structure make direct numerical simulation extremely demanding. Particle-In-Cell (PIC) methods remain the standard for ion-scale studies, yet their memory cost and intrinsic noise hinder accurate electron-scale simulations. In this paper, we introduce an alternative method based on an iterative- in-time approximation of characteristics. The approach reconstructs the phase-space dynamics from the time history of the electromagnetic fields and the initial distribution functions, enabling extremely high effective resolution far below the phase-space grid scale without storing or advecting high-dimensional data. Earlier work demonstrated this capability for the multi-species electrostatic Vlasov system. Here we discuss an extension of the method to the full Vlasov–Maxwell equations using a Hamiltonian splitting to advance the solution in a structure-preserving way while retaining the reduced memory footprint.

## 1 Introduction

When a plasma becomes too hot or rarefied, the velocity distribution no longer remains Maxwellian, i.e., particles can substantially deviate from thermal equilibrium, and thus the fundamental assumption of a fluid-description such as Magnetohydrodynamics (MHD) or related moment methods breaks down. Then the plasma evolution has to be modelled via kinetic theory, i.e., including the velocity distributions in the model, which results in the Vlasov equation[1]

$$\partial_t f^\alpha + \boldsymbol{v} \cdot \nabla_{\boldsymbol{x}} f^\alpha + \tfrac{q_\alpha}{m_\alpha} (\boldsymbol{E} + \boldsymbol{v} \times \boldsymbol{B}) \cdot \nabla_{\boldsymbol{v}} f^\alpha = 0. \tag{1}$$

where $f^\alpha$ is the probability distribution of the species $\alpha$ in the up to six-dimensional phase-space. $\boldsymbol{E}$ and $\boldsymbol{B}$ are the (self-induced) electro-magnetic fields acting upon the species $\alpha$ computed via Maxwell's equations

$$\nabla \times \boldsymbol{E} + \partial_t \boldsymbol{B} = 0, \qquad \nabla \times \boldsymbol{B} - \partial_t \boldsymbol{E} = \boldsymbol{j}, \qquad (2)$$

$$\nabla \cdot \boldsymbol{E} = \rho, \qquad \nabla \cdot \boldsymbol{B} = 0. \qquad (3)$$

The right hand side of the Maxwell's equations is computed from the phase-space distribution functions $f^\alpha$:

$$\rho(t, \boldsymbol{x}) = \sum_\alpha \int_{\mathbb{R}^d} f^\alpha(t, \boldsymbol{x}, \boldsymbol{v}) \mathrm{d}\boldsymbol{v}, \qquad \boldsymbol{j}(t, \boldsymbol{x}) = \sum_\alpha \int_{\mathbb{R}^d} \boldsymbol{v} f^\alpha(t, \boldsymbol{x}, \boldsymbol{v}) \mathrm{d}\boldsymbol{v}. \qquad (4)$$

With this coupling equations (1) – (4) become the (non-linear) Vlasov–Maxwell system. We use the normalized version of the Vlasov–Maxwell system in rationalized Gaussian units following Morrison,[2] details can be found in appendix A.

Solving the Vlasov–Maxwell system is computationally demanding because the dynamics evolve in six-dimensional phase space, so even low-resolution simulations are expensive for methods that directly discretize the distribution functions. In many astrophysical environments, such as the solar wind, plasmas are effectively collisionless and thus develop fine multi-scale structures that must be resolved to capture kinetic instabilities and energy dissipation mechanisms.[1,3–5] Combined with the large electron-ion mass ratio and the disparity between system and particle dynamics scales, this makes the Vlasov system a strongly multi-scale problem,[3] where high resolution is required but usually unaffordable.

Numerical methods for the Vlasov equation are often grouped into Eulerian, Semi-Lagrangian, and Lagrangian (particle) schemes. Eulerian and Semi-Lagrangian solvers represent the distribution function on a phase-space grid and have been successfully applied in kinetic plasma and space-physics simulations.[6–19] However, the six-dimensional grid incurs extreme storage and communication costs, especially when dealing with highly heterogeneous flows or complex geometries, and introduces artificial diffusion that conflicts with the conservative, collisionless nature of the Vlasov dynamics. Particle-In-Cell (PIC) methods instead represent $f$ by marker particles evolved along characteristics, coupled to electromagnetic fields on a spatial grid.[20–39] Particles automatically adapt to the flow dynamics, making them well-suited for highly heterogenuous flow, and their unstructured nature simplifies parallelization, but their inherent noise requires the use of many particles for reliable results, hence making high-fidelity simulations very costly. A detailed overview of numerical solvers for the Vlasov equation can be found in e.g. the reviews by Filbet and Sonnendrücker[40] or, more recently, Palmroth et al.[18]

Following the idea of Semi-Lagrangian schemes, a new class of schemes was recently introduced, which shifts the focus to approximation of the flow map instead of the distribution function. The two main approaches in this direction are the Characteristic Mapping Method (CMM) introduced by Krah et al.[41,42] and the Numerical Flow Iteration (NuFI) introduced by Wilhelm et al.[43,44] Solving for the flow map naturally preserves the solution structure of the Vlasov–Maxwell system, and, if using time-stepping based upon a Hamiltonian splitting, this also leads to optimal conservation properties of the numerical scheme. While CMM directly stores the flow-map, exploiting the sub-group property of the flow to compress the data, NuFI approximates the flow on-the-fly via an iterative backwards-in-time scheme built upon symplectic time-integration. The NuFI approach allows for high-fidelity simulations while keeping the computational costs low as only the time-evolution of the lower-dimensional electro-magnetic fields has to be stored instead of the high-dimensional distribution function data. Additionally, because NuFI only looks at the characteristics instead of a fixed discretization of the distribution function in phase-space, the method has extreme sub-grid resolution properties and, on the other hand, allows easy introduction of adaptivity and handling of boundary conditions, while remaining computationally efficient.[42,45,46]

In this paper, we want to present an extension of the NuFI approach to the full Vlasov–Maxwell system in section 2 and show case the high-fidelity of NuFI for resolving fine structures in phase-space on a $2x3v$ test benchmark in section 3.

## 2 The Numerical Flow Iteration for the Vlasov–Maxwell system

For the sake of brevity we will focus on the NuFI scheme for the Vlasov–Maxwell system. Detailed derivation and discussion of the electro-static case can be found in literature.[43,45,46]

*2.1 Hamiltonian of the Vlasov–Maxwell system*

The Vlasov–Maxwell system is a Hamiltonian system with[2,47,48]

$$\begin{aligned}\mathcal{H}_{VM} &= \mathcal{H}_{\boldsymbol{E}} &&+ \mathcal{H}_{\boldsymbol{B}} &&+ \mathcal{H}_f \\ &= \tfrac{1}{2}\int_{\mathbb{R}^d} \boldsymbol{E}^2 \mathrm{d}\boldsymbol{x} &&+ \tfrac{1}{2}\int_{\mathbb{R}^d} \boldsymbol{B}^2 \mathrm{d}\boldsymbol{x} &&+ \tfrac{1}{2}\sum_\alpha \int_{\mathbb{R}^d\times\mathbb{R}^d} f^\alpha \mathrm{d}\boldsymbol{x}\mathrm{d}\boldsymbol{v}.\end{aligned} \qquad (5)$$

The three terms $\mathcal{H}_{\boldsymbol{E}}, \mathcal{H}_{\boldsymbol{B}}$ and $\mathcal{H}_f$ correspond to the electric, magnetic and kinetic energy respectively. Each of the sub-Hamiltonian admits a corresponding linear advection equation:[48] The equations associated to the electric field Hamiltonian $\mathcal{H}_{\boldsymbol{E}}$ are

$$\partial_t f^\alpha + \boldsymbol{E} \cdot \nabla_{\boldsymbol{v}} f^\alpha = 0, \qquad (6)$$

$$\partial_t \boldsymbol{E} = 0, \qquad \partial_t \boldsymbol{B} = -\nabla_{\boldsymbol{x}} \times \boldsymbol{E}. \qquad (7)$$

The equations associated to the electric field Hamiltonian $\mathcal{H}_{\boldsymbol{E}}$ are

$$\partial_t f^\alpha + (\boldsymbol{v}\times\boldsymbol{B}(\boldsymbol{x}))\cdot\nabla_{\boldsymbol{v}} f^\alpha = 0, \qquad (8)$$

$$\partial_t \boldsymbol{E} = \nabla_{\boldsymbol{x}} \times \boldsymbol{B}, \qquad \partial_t \boldsymbol{B} = 0. \qquad (9)$$

And finally the equations associated to the electric field Hamiltonian $\mathcal{H}_f$ are

$$\partial_t f^\alpha + \boldsymbol{v}\nabla_{\boldsymbol{x}} f^\alpha = 0, \qquad (10)$$

$$\partial_t \boldsymbol{E} = -\boldsymbol{j}, \qquad \partial_t \boldsymbol{B} = 0. \qquad (11)$$

Because the above sub-systems are linear advection equations they all admit analytic solutions, which can be used to construct an approximate numerical solution via an operator-splitting approach. Note that this is the same principle behind classical time integration schemes like the *Leapfrog* commonly used in PIC codes. The analytical solution to (6) – (7) is

$$f^\alpha(t,\boldsymbol{x},\boldsymbol{v}) = f_0^\alpha\left(\boldsymbol{x},\boldsymbol{v} - t\boldsymbol{E}_0\left(\boldsymbol{x}\right)\right), \qquad (12)$$

$$\boldsymbol{E}(t,\boldsymbol{x}) = \boldsymbol{E}_0(\boldsymbol{x}), \qquad \boldsymbol{B}(t,\boldsymbol{x}) = \boldsymbol{B}_0(\boldsymbol{x}) - t\nabla_{\boldsymbol{x}}\times\boldsymbol{E}_0(\boldsymbol{x}), \qquad (13)$$

provided the initial data $\boldsymbol{E}_0, \boldsymbol{B}_0$ and $f_0$ at time $t = 0$. Similarly for (8) – (9) we get

$$f^\alpha(t,\boldsymbol{x},\boldsymbol{v}) = f_0^\alpha\left(\boldsymbol{x}, \exp\left(-\mathbb{J}_{\boldsymbol{B}} t\right)\boldsymbol{v}\right), \qquad (14)$$

$$\boldsymbol{E}(t,\boldsymbol{x}) = \boldsymbol{E}_0(\boldsymbol{x}) + t\nabla_{\boldsymbol{x}}\times\boldsymbol{B}_0(\boldsymbol{x}), \qquad \boldsymbol{B}(t,\boldsymbol{x}) = \boldsymbol{B}_0(\boldsymbol{x}), \qquad (15)$$

where $\mathbb{J}_{\boldsymbol{B}}$ denotes the rotation matrix with $\boldsymbol{B}$, and for (10) – (11)

$$f^\alpha(t, \boldsymbol{x}, \boldsymbol{v}) = f_0^\alpha\left(\boldsymbol{x} - t\boldsymbol{v}, \boldsymbol{v}\right), \tag{16}$$

$$\boldsymbol{E}(t, \boldsymbol{x}) = \boldsymbol{E}_0(\boldsymbol{x}) + \int_0^t \sum_\alpha \left(\int_{\mathbb{R}^3} \boldsymbol{v} f_0^\alpha\left(\boldsymbol{x} - s\boldsymbol{v}, \boldsymbol{v}\right) \mathrm{d}\boldsymbol{v}\right) \mathrm{d}s, \qquad \boldsymbol{B}(t, \boldsymbol{x}) = \boldsymbol{B}_0(\boldsymbol{x}). \tag{17}$$

From these analytic solution a numerical approximation can be constructed through choosing an operator splitting. We decided for first order Lie splitting for its simplicity, though in principle higher order is also possible. In short, there are two options to construct a scheme from the above: If one decides to use the Hamiltonian splitting for both the Vlasov advection and solving the field equations (called *NuFI-Ham*), this is highly conservative, because it fully preserves the Hamiltonian structure, however, this naturally introduces an interdependency between the velocity space discretization and time-integration for the Maxwell solver (due to the integral to obtain the current density). This makes the CFL condition introduced through Maxwell's equations depend on $\Delta \boldsymbol{v}$ in addition to $\Delta \boldsymbol{x}$, which often severely restricts the allowed time step size similar to purely Eulerian approaches.[11,44]

Instead, if choosing to only use the Hamiltonian splitting for the Vlasov advection and another scheme for the Maxwell solver,[44,48] this eliminates the $\Delta \boldsymbol{v}$ dependency in the CFL condition at the cost of slightly worse conservation properties, albeit still significantly better than that of classical Semi-Lagrangian approaches. For this work we decided for a coupling to a *predictor-corrector* Maxwell solver[21] resulting in the so-called *NuFI-PC* scheme.

The NuFI-update formula is

$$f(t_{n+1}, \boldsymbol{x}, \boldsymbol{v}) = f(t_n, \tilde{\boldsymbol{x}}, \tilde{\boldsymbol{v}}(\tilde{\boldsymbol{x}})), \tag{18}$$

where we denote $\tilde{\boldsymbol{x}} = \boldsymbol{x} - \Delta \boldsymbol{v}$ and $\tilde{\boldsymbol{v}}(\boldsymbol{x}) = \exp(-\frac{q_\alpha}{m_\alpha}\Delta t J_{\boldsymbol{B}_0 - \Delta t \nabla_{\boldsymbol{x}} \times \boldsymbol{E}_0(\boldsymbol{x})})\boldsymbol{v} - \Delta t \frac{q_\alpha}{m_\alpha}\boldsymbol{E}_0(\boldsymbol{x})$.

We discretize $\boldsymbol{E}$ and $\boldsymbol{B}$ on a uniform in space but staggered in time grid through

$$\boldsymbol{E}^{n+1} = \boldsymbol{E}^n + \Delta t \nabla_{\boldsymbol{x}} \times \boldsymbol{B}^{n+1/2} - \tfrac{\Delta t}{2}(\boldsymbol{j}^n + \boldsymbol{j}^{n+1}), \tag{19}$$

$$\boldsymbol{B}^{n+1/2} = \boldsymbol{B}_{n-1/2} - \Delta t \nabla_{\boldsymbol{x}} \times \boldsymbol{E}^n. \tag{20}$$

The thus obtained values are then interpolated on the spatial grid using B-Splines. Because we require $\boldsymbol{B}^n$ for (18), we also compute and interpolate

$$\boldsymbol{B}^n = \tfrac{1}{2}(\boldsymbol{B}^{n+1/2} + \boldsymbol{B}^{n-1/2}). \tag{21}$$

Furthermore, if we only prescribe $\boldsymbol{B}_0$ then $\boldsymbol{B}_{-1/2}$ can be predicted via

$$\boldsymbol{B}^{-1/2} = \boldsymbol{B}_0 + \tfrac{\Delta t}{2}\nabla_{\boldsymbol{x}} \times \boldsymbol{E}_0. \tag{22}$$

This leaves us with the following NuFI-PC algorithm

**Algorithm 1** Numerical Flow Iteration for the (multi-species) Vlasov–Maxwell system coupled with Predictor-Corrector. (NuFI-PC)

**function** NuFI($f_0^\alpha$, $\boldsymbol{B}_0$, $\boldsymbol{E}_0$, $N_t$, $\Delta t$, $(N_{x_i}^\alpha)_{i,\alpha}$, $(N_{v_i}^\alpha)_{i,\alpha}$, $(v_{\min}^{i,\alpha})_{i,\alpha}$, $(v_{\max}^{i,\alpha})_{i,\alpha}$ )
    Interpolate $\boldsymbol{E}_0$ and $\boldsymbol{B}_0$ on spatial grid.
    Compute and interpolate $\boldsymbol{B}_{-1/2}$ via (22).
    Compute $\boldsymbol{j}_0$ on spatial grid using (18) and a midpoint rule.
    Compute and interpolate $\boldsymbol{B}_{1/2}$ using (20).
    **for** $n = 1, ..., N_t$ **do**
        Compute $\boldsymbol{j}_n$ on spatial grid using (18) and a midpoint rule.
        Compute and interpolate $\boldsymbol{E}^n$ using (19).
        Compute and interpolate $\boldsymbol{B}_{n+1/2}$ using (20).
        Compute and interpolate $\boldsymbol{B}^n$ using (21).
        $\boldsymbol{j}_{n-1} = \boldsymbol{j}_n$.
    **end for**
**end function**

A detailed derivation of the NuFI-PC (and NuFI-Ham) scheme can be found in the recent publication by Wilhelm et al.[44]

*2.2 Reducing the Computational Complexity of NuFI*

The major drawback of the NuFI approach – both in the electro-static and electro–magnetic case – is the increased computational complexity. The evaluation of $f$ is done on-the-fly through the aforementioned backwards-in-time iteration, which however implies that to evaluate $f(n)$ one has to iterate $n$ time-steps backwards in time resulting in linear complexity for a single evaluation. For a full simulation with a total of $n_t$ time steps the computational complexity thus becomes $\mathcal{O}(n_t^2)$. Therefore simulating long time periods becomes prohibitively expensive with the pure NuFI, but this can be remedied through restarting the simulation at *checkpoints* of either the distribution function or the flow map, which effectively reduces the computational complexity of a full simulation back to linear.[42,45,49]

In the current (prototype) implementation of NuFI we employ restarts via storing the distribution function a uniform grid and using linear interpolation to evaluate it, which effectively makes it similar to a linear Semi-Lagrangian approach with sub-cycling for all species, however, with substantially less frequent need to store the intermediate results. While simplistic this form of storage is computationally efficient and, more importantly, allows for straight-forward low rank compression.[49,50]

## 3 Simulation of a 2x3v Beam Instability

In this section we want show case the fidelity of the NuFI approach with focus on the resolution of the phase-space velocity distribution. To this end we consider a two beam instability in 2 spatial and 3 velocity dimensions. On time scales we are interested in to simulate, we can safely assume a uniform ion background. The initial electron distribution function is

$$f_0(x,y,u,v,w) = \frac{1+\alpha \cos(x)\cos(y)}{2(2\pi v_{\text{th}}^2)^{3/2}} \left( \exp\left(-\frac{(u-v_d)^2}{2v_{\text{th}}^2}\right) + \exp\left(-\frac{(u+v_d)^2}{2v_{\text{th}}^2}\right) \right) \exp\left(-\frac{v^2+w^2}{2v_{\text{th}}^2}\right), \quad (23)$$

where we set $v_{\text{th}} = 1$, $\alpha = 0.01$ and $v_d = 2$. The initial electric field can be computed from the Gauss law resulting in $\boldsymbol{E}_0(x,y) = \frac{\alpha}{2}(\sin(x)\cos(y), \cos(x)\sin(y), 0)^T$. Furthermore, we prescribe an initial magnetic perturbation to also trigger magnetic modes $\boldsymbol{B}_0(x,y) = B_0(0, 0, \cos(x)\cos(y))^T$ with $B_0 = 0.1$. The physical domain is set to $[0, 2\pi] \times [0, 2\pi]$ and the velocity space is truncated to $[-30, 30] \times [-30, 30] \times [-10, 10]$. The fields interpolation is using B-Splines of 4th order and we restart the simulation with a linear interpolant on an uniform grid coinciding with the grid used to evaluate $\boldsymbol{j}$, which is set to $16 \times 16$ in space and $48 \times 48 \times 16$ in velocity. The restarts are performed every $n_t^r = 20$ time steps. The final time was set to $T = 300$. The total simulation[1] time was $\approx 2600s$ on 1 node of the Tier-1 VSC-cluster *Hortense*.[2]

In figure 1 we show the evolution of the field energies. After a short initial phase of periodic oscillation on a roughly constant energy level, the electric energy grows (approximately) linearly after $t \approx 20$ until saturates at $t \approx 150$. The magnetic energy oscillates on about the its initial level until $t \approx 30$ after which it slowly decreases by an order of magnitude until $t \approx 110$. After $t \approx 110$ there is a short phase in which the magnetic energy also grows linearly by approximately 2 orders of magnitude. After $t \approx 120$ both the electric and magnetic energy reach a saturation level and remain constant until the end of the simulation. In figure 1b, we see that the dominating components in the dynamics are the $x$- and $y$-component of the electric and the $z$-component of the magnetic field. Until roughly $t \approx 130$ the other components remain constantly at 0 (below machine precision) and afterward they increase to a constant level around $10^{-5}$.

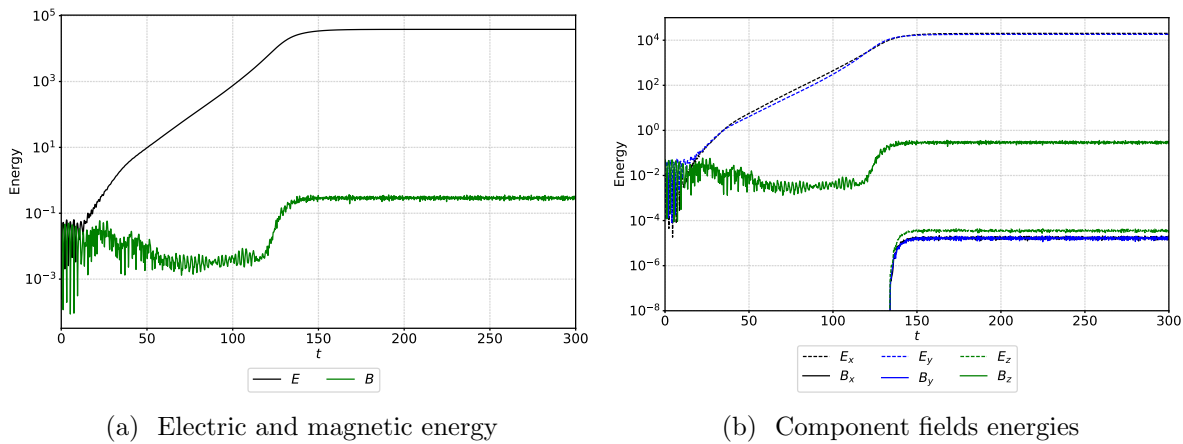

(a)  Electric and magnetic energy

(b)  Component fields energies

Figure 1: The figure shows the evolution of electromagnetic field energies. Left we show the total electric and magnetic energy, while right the energies of the single components are shown.

In figure 2, we first look at the distribution in space (at $\boldsymbol{v} = 0$). The initial checkerboard pattern of the density is transitioned into "island-like" structures along the beam-velocity direction $x$. In figure 3, we look at the distribution function in $x$-$u$ direction at $y = \pi$ and 0 in the other velocity directions. Initially we observe mixing between the two beams, similarly to the electro-static two stream instability, however, after the first vortex forms around $t \approx 100$ we observe a rapid blow-up of the velocity support. For the late stage of the instability the distribution function is highly filamented with small phase-space vortices, which however are no longer necessarily aligned with the initial $x$ direction along which the instability formed. In figure 4, we now also look at $f$ in the first two components of the velocity space at $x = y = \pi$ and $w = 0$. While at first we observe a merging between the two beams, the distribution function becomes increasingly filamented over time and we also clearly observe the velocity support blowing up for late times.

---

[1] The code used for the simulations can be found on the NuFI GitHub on *tensor_svd_approx* branch.
[2] Each node has $2 \times 64$-cores AMD Epyc 7763 CPUs (2.45 GHz) with 238 GiB RAM.

The main advantage of NuFI over other Vlasov solvers is its extreme subgrid resolution property also referred to as *Zoom property*:[41–44] Even though this simulation runs with a relatively low computational resolution, the dynamics are still captured well and we can see details on subgrid-scales, beyond what a classical Vlasov solver can achieve. That is because classic Vlasov solvers have to discretize the distribution function with a given resolution each time step and thereby can only preserve information about the solution to the chosen resolution (possibly also taking discretization order into account), however, this is not the case for NuFI, which mainly relies upon storing the smoother fields and doesn't truncate the distribution function data as frequently (with the restart frequency), which leads to much more accurate representation of the flow.

Already in the full views of the distribution function, figures 2-4, the level of detail is high. Keeping the same computational resolution, the plotting resolution is increased to 512 in each displayed direction, which allows us to visualize details beyond the very coarse computational grid. However, in figure 5 we take this a step further: Because NuFI is not restricted to a certain phase-space resolution we can *zoom* arbitrary far into the distribution function. In figure 5a we zoom into a $4 \times 4$ spatial cell block of $f(150, x, y, 0, 0, 0)$ to look in detail at one of the "plasma blobs". The filamentation structure can be well seen far at scales far below the actual coarse grid. In figure 5b we zoom into a single velocity space cell of $f(150, \pi, \pi, u, v, 0)$ and also here observe details at scales orders of magnitude smaller than the grid-scale. Note that the "grainy spots" in the plots 5a and 5b are not numerical noise as would be expected for e.g. a PIC simulation, but instead are a lack of plotting resolution, i.e., could be clearly resolved by further increasing the resolution of the plot.[43]

**Remark 3.1** *In figures 2c, 3c, 4c and 5, we introduced a logarithmic color scale for better visibility of the fine scale structures. Zero entries are white and, in post-processing, values smaller than the minimum value on the color bar are shown in the same color as the minimum of the color bar.*

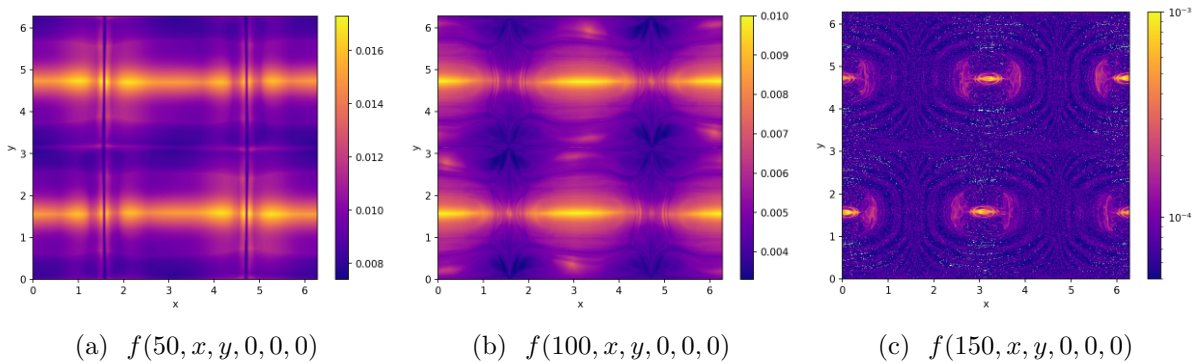

(a) $f(50, x, y, 0, 0, 0)$  (b) $f(100, x, y, 0, 0, 0)$  (c) $f(150, x, y, 0, 0, 0)$

Figure 2: Slice of the distribution function $f(t, x, y, 0, 0, 0)$ shown at times $t = 50$ (left), 100 (middle) and 150 (right).

## 4 Discussion and outlook

In this work we presented an extension of the Numerical Flow Iteration (NuFI) approach to the electro-magnetic Vlasov–Maxwell system called Numerical Flow Iteration Predictor-Corrector (NuFI-PC) scheme. Its properties as well as practical implementation have been discussed. We

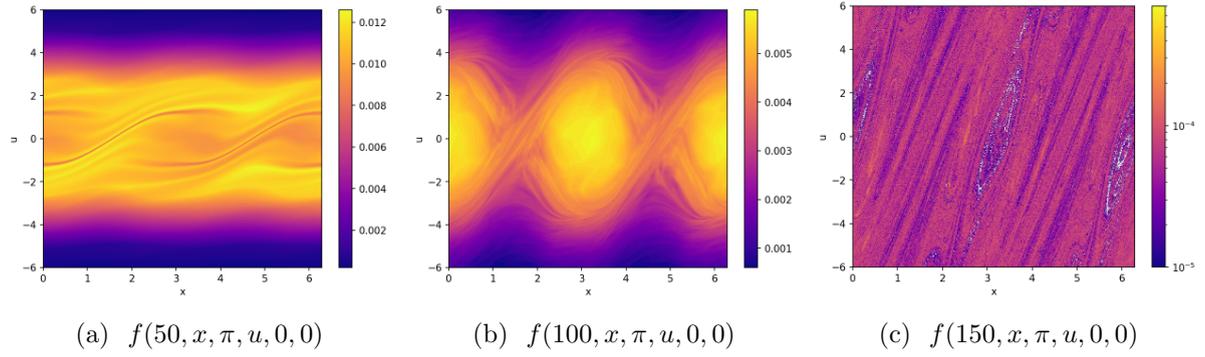

(a) $f(50, x, \pi, u, 0, 0)$  (b) $f(100, x, \pi, u, 0, 0)$  (c) $f(150, x, \pi, u, 0, 0)$

Figure 3: Slice of the distribution function $f(t, x, \pi, u, 0, 0)$ shown at times $t = 50$ (left), 100 (middle) and 150 (right).

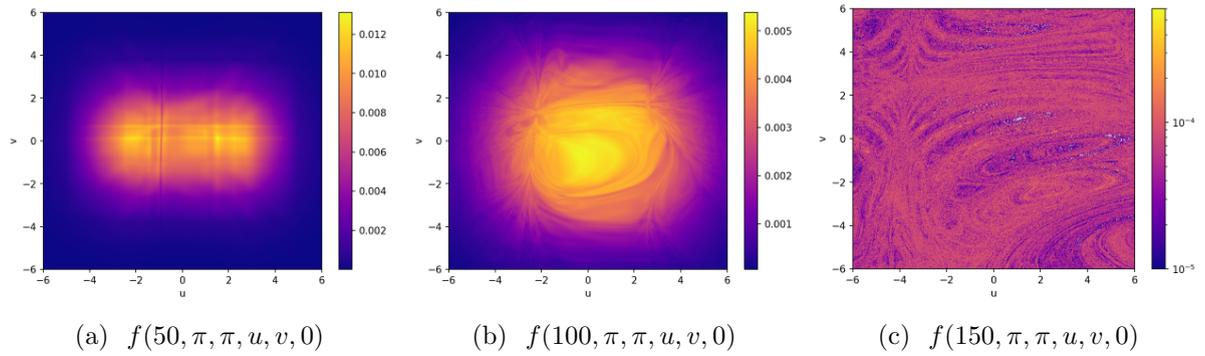

(a) $f(50, \pi, \pi, u, v, 0)$  (b) $f(100, \pi, \pi, u, v, 0)$  (c) $f(150, \pi, \pi, u, v, 0)$

Figure 4: Slice of the distribution function $f(t, \pi, \pi, u, v, 0)$ shown at times $t = 50$ (left), 100 (middle) and 150 (right).

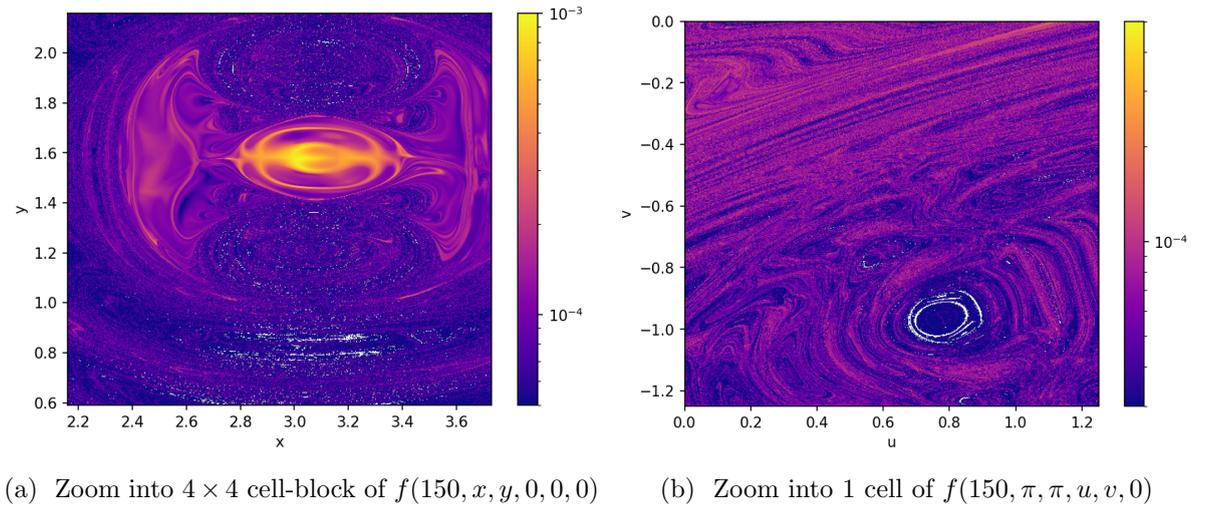

(a) Zoom into $4 \times 4$ cell-block of $f(150, x, y, 0, 0, 0)$  (b) Zoom into 1 cell of $f(150, \pi, \pi, u, v, 0)$

Figure 5: Here we zoomed into the distribution function at $t = 150$ taken from figures 2c and 4c. The left figure shows a zoom into a $4 \times 4$ spatial cell block of $f(150, x, y, 0, 0, 0)$ and the right figure shows a zoom into 1 velocity cell of $f(150, \pi, \pi, u, v, 0)$.

showcased the accuracy and, in particular, the extreme subgrid resolution property of NuFI through a simulation of a beam instability in five-dimensional phase-space. In on-going and future work we want to look into simulations of fully six-dimensional, multi-species kinetic plasma dynamics, e.g. of beam-driven instabilities,[51,52] down to the electron scale. Due to the strong subgrid resolution of NuFI, we expect that good results can be achieved even with relatively coarse resolutions. Furthermore, we actively investigate how the restart procedure can be made more efficient and accurate. While the current procedure is accurate enough on the relatively short time scales we considered here, the dissipative nature of linear interpolation leads to a significant loss of accuracy for long simulation periods.

**Acknowledgments**
This work has received funding from the European High Performance Computing Joint Undertaking (JU) and Belgium, Czech Republic, France, Germany, Greece, Italy, Norway, and Spain under grant agreement No 101093441. Views and opinions expressed are however those of the author(s) only and do not necessarily reflect those of the European Union or the European High Performance Computing Joint Undertaking (JU) and Belgium, Czech Republic, France, Germany, Greece, Italy, Norway, and Spain. Neither the European Union nor the granting authority can be held responsible for them. F.B. acknowledges support from the FED-tWIN programme (profile Prf-2020-004, project "ENERGY") issued by BELSPO, and from the FWO Junior Research Project G020224N granted by the Research Foundation – Flanders (FWO). The resources and services used in this work were provided by the VSC (Flemish Supercomputer Center), funded by the Research Foundation - Flanders (FWO) and the Flemish Government.

## A  Normalization of the multi-species Vlasov–Maxwell system

We choose to normalize the Vlasov–Maxwell system to the electron scales, i.e., we choose as reference the electron charge $q_e = e$, electron mass $m_e$ and the electron number density $n_0$. The electron plasma frequency is $\omega_{p,e}^2 = \frac{4\pi n_0 e^2}{m_e}$. Hence we choose as a reference scale for time $t_0 = \omega_{p,e}^{-1}$, spatial dimension $L_0 = \frac{c}{\omega_{p,e}}$ and velocity $V_0 = c$. Consequentially we choose for the fields $E_0 = \frac{m_e c \omega_{p,e}}{e}$, $B_0 = E_0$, $f_0 = \frac{n_0}{V_0^3} = \frac{n_0}{c^3}$. Finally the charge and current density are scaled to $\rho_0 = en_0$ and $J_0 = ecn_0$ respectively. Then our dimensionless variables used in (1) – (??) are $t' = \frac{t}{t_0}, x' = \frac{x}{L_0}, v' = \frac{v}{V_0}, f'_s = \frac{f_s}{f_0}, E' = \frac{E}{E_0}, B' = \frac{B}{B_0}, \rho' = \frac{\rho}{\rho_0}, J' = \frac{J}{J_0}$. For the sake of readability we drop the primes in our notation.